\documentclass[article]{jss}

\usepackage{orcidlink,thumbpdf,lmodern, booktabs, tikz, bm, amsmath, amssymb}

\newcommand{\class}[1]{`\code{#1}'}

\def\checkmark{\tikz\fill[scale=0.4](0,.35) -- (.25,0) -- (1,.7) -- (.25,.15) -- cycle;} 
\newcommand{\xmark}{%
\tikz[scale=0.23] {
    \draw[line width=0.7,line cap=round] (0,0) to [bend left=6] (1,1);
    \draw[line width=0.7,line cap=round] (0.2,0.95) to [bend right=3] (0.8,0.05);
}}

\author{Alvaro Mendez-Civieta~\orcidlink{0000-0003-2044-4170}\\University of Columbia\\uc3m-Santander Big Data Institute
   \And M. Carmen Aguilera-Morillo~\orcidlink{0000-0003-1027-9773}\\Universitat Politècnica de València
   \AND Rosa E. Lillo~\orcidlink{0000-0003-0802-4691}\\uc3m-Santander Big Data Institute\\Universidad Carlos III de Madrid. }
\Plainauthor{Alvaro Mendez-Civieta, M. Carmen Aguilera-Morillo, Rosa E. Lillo}

\title{\pkg{asgl}: A \proglang{Python} Package for Penalized Quantile, Linear and Logistic Regression}
\Plaintitle{asgl: A Python Package for Penalized Quantile, Linear and Logistic Regression}

\Abstract{
\pkg{asgl} is an open-source \proglang{Python} package that offers a robust and versatile framework for fitting a variety of regression models including linear, logistic, and, notably, quantile regression. It implements a comprehensive suite of penalization techniques such as lasso, ridge, group lasso, sparse group lasso, elastic net, and their adaptive variants. A key contribution of \pkg{asgl} is its extensive support for adaptive penalizations, critically offering a range of built-in methodologies for estimating the necessary adaptive weights as proposed by Mendez-Civieta et al. (2020). This feature addresses a significant practical challenge -the weight estimation process -in applying advanced adaptive methods, especially in high-dimensional settings, and is largely absent from other packages. Furthermore, \pkg{asgl} offers penalized quantile regression, a less commonly available feature in statistical software.  The primary class, \code{Regressor}, ensures seamless integration with the \pkg{scikit-learn} ecosystem, facilitating straightforward model evaluation and hyperparameter optimization. \pkg{asgl} has demonstrated utility in variable selection and prediction tasks across both low- and high-dimensional data, positioning it as a comprehensive tool for modern statistical modeling.
}

\Keywords{regression, regularization, variable-selection, high-dimension, adaptive penalization, quantile regression, \proglang{python}}
\Plainkeywords{regression, regularization, variable-selection, high-dimension, adaptive penalization, quantile regression, python}

\Address{
  Alvaro Mendez-Civieta\\
  Department of Biostatistics\\
  Columbia University in the City of New York\\
  722 West 168th Street\\
  10032, New York, NY, USA\\
  E-mail: \email{am5490@cumc.columbia.edu}\\
}

\begin{document}


\section[Introduction]{Introduction}\label{sec:introduction}

Linear, logistic, and quantile regression are foundational tools in statistical modeling. While linear and logistic regression are extensively used for continuous and categorical outcomes respectively, quantile regression, since the seminal work by \citet{Koenker1978}, offers distinct advantages such as robustness to outliers, skewness, and the ability to model different parts of the conditional distribution, making it invaluable for heteroscedastic datasets.

In contemporary data analysis, particularly in fields like genetics \citep{Simon2013}, finance \citep{Rapach2013}, and pattern recognition \citep{Wright2010}, datasets are often high-dimensional, with the number of predictors $p$ exceeding the number of observations $n$. Such scenarios necessitate penalized regression methods to prevent overfitting and select relevant variables. Various penalization techniques have been developed to address these challenges. Ridge regression \citep{hoerl_ridge_1970} applies an $\ell_2$ penalty to handle multicollinearity but does not perform variable selection. The lasso (least absolute shrinkage and selection operator) by \citet{Tibshirani1996} uses an $\ell_1$ penalty to achieve sparse solutions and automatic variable selection. The elastic net \citep{Zou2005} combines $\ell_1$ and $\ell_2$ penalties. Group-wise penalties, such as the group lasso \citep{Yuan2006} and sparse group lasso \citep{friedman_note_2010, simon_sparse-group_2013}, extend these ideas to select groups or individual variables within groups.

Despite their effectiveness, these standard penalizations can yield biased estimates due to their constant penalization rates. \citet{Zou2006a} addressed this by introducing the adaptive lasso, incorporating variable-specific weights into the penalty term to mitigate bias and improve variable selection consistency. Usually, these weights are derived from initial estimates (e.g., from unpenalized models), limiting adaptive methods to low-dimensional settings. \citet{Mendez-Civieta2020} extended this by defining adaptive sparse group lasso (ASGL) estimators and proposing novel weight calculation methods using Principal Component Analysis (PCA) or Partial Least Squares (PLS) among others, effective in both low- and high-dimensional contexts.

The \pkg{asgl} Python package is designed to address the need for a comprehensive and user-friendly tool for these advanced regression techniques. Built upon \pkg{cvxpy} \citep{Diamond2016}, \pkg{asgl} offers robust solvers for convex optimization problems and provides a unified framework for solving linear, logistic, and quantile regression models using a wide array of penalizations: ridge, lasso, group lasso, sparse group lasso, elastic net, and, crucially, the adaptive versions of all these methods. The core contributions of \pkg{asgl} are: (1) a versatile platform for multiple model types (linear, logistic, quantile); (2) an extensive library of penalizations, with a strong emphasis on their adaptive variants; (3) the implementation of several methodologies for estimating adaptive weights, a critical step for the practical application of adaptive penalizations, especially in high-dimensional scenarios where unpenalized estimates for weights are infeasible; (4) robust support for penalized quantile regression; and (5) full integration with the \pkg{scikit-learn} \citep{scikit-learn} ecosystem, enabling easy use of tools like \code{GridSearchCV} and \code{RandomizedSearchCV} for hyperparameter tuning and allowing users to substitute \pkg{asgl} estimators into existing pipelines with minimal code changes.

While several packages offer penalized regression, \pkg{asgl} provides unique advantages. For standard non-adaptive penalizations in \proglang{Python}, \pkg{scikit-learn} is a popular choice for lasso, ridge, and elastic net. \pkg{groupyr} offers group lasso and sparse group lasso for linear models. In \proglang{R}, \pkg{glmnet} \citep{Friedman2010b} is widely used for lasso, elastic net, and their adaptive forms (for linear and logistic regression), and \pkg{sparsegl} \citep{liang_sparsegl_2024} handles lasso, group lasso, and sparse group lasso. For quantile regression, \pkg{conquer} \citep{he_smoothed_2023} in \proglang{R} and \pkg{scikit-learn} (for lasso) offer some penalized options. By adjusting hyperparameters, users can extend these packages' functionalities: for instance, setting specific hyperparameters within group lasso or elastic net penalizations can produce ridge models, and modifying sparse group lasso settings can yield ridge, lasso, group lasso, and elastic net models. However, the landscape for adaptive penalizations, particularly with integrated weight estimation, is less developed. For instance, while \pkg{glmnet} supports adaptive lasso/elastic net, it requires users to provide the adaptive weights, a non-trivial task, especially in high dimensions. \pkg{asgl} fills this gap by offering not only a wide array of adaptive penalizations (including adaptive group lasso and adaptive sparse group lasso, which are not commonly found) but also multiple built-in techniques to estimate these crucial weights automatically. Furthermore, complex models like the adaptive sparse group lasso cannot be trivially implemented by scaling covariates (a trick sometimes used for adaptive lasso) and necessitate a dedicated formulation and solver, which \pkg{asgl} provides across all its supported regression types. Table~\ref{tab:comparison} provides an overview of \pkg{asgl}'s capabilities relative to other prominent packages.

The development of a dedicated package like \pkg{asgl} offers significant benefits over implementing these complex models from scratch or relying on piecemeal solutions. It reduces the considerable implementation burden, particularly for advanced methods like adaptive penalizations and the various adaptive weight estimation strategies. A well-tested and documented package promotes reproducibility, reusability, and accessibility for a broader range of users who may not have the expertise or time to develop these methods independently. The seamless integration with \pkg{scikit-learn} further lowers the barrier to entry for adopting these powerful adaptive techniques by allowing users to leverage familiar workflows for hyperparameter tuning and model validation. Thus, \pkg{asgl} aims to make sophisticated penalized and adaptive regression methods more readily available and practical for the statistics and machine learning communities.

The remainder of this paper is organized as follows. Section~\ref{sec:theory} details the theoretical background of the regression models and penalizations implemented in \pkg{asgl}, with a particular focus on adaptive methods and weight calculation techniques. Section~\ref{sec:python_implementation} describes the \proglang{Python} implementation, focusing on the main \code{Regressor} class and its functionalities. Section~\ref{sec:examples} provides practical usage examples to illustrate the capabilities of the package. Section~\ref{sec:computational_details} discusses some practical details. Section~\ref{sec:limitations} discusses the limitations of the current version of \pkg{asgl}. Finally, Section~\ref{sec:conclusions} summarizes the contributions and concludes the paper.

\begin{table}[htb]
\centering
    \caption{Comparison of \pkg{asgl} against other alternatives in \proglang{Python} and \proglang{R}. Adaptive weight estimation refers to built-in methodologies.}
    \label{tab:comparison}
    \resizebox{\textwidth}{!}{
    \begin{tabular}{lllllll}
    \toprule
                                & \pkg{asgl}                        & \pkg{sparsegl}                    & \pkg{glmnet}                      & \pkg{conquer}                     & \pkg{groupyr}                     & \pkg{scikit-learn}                \\ \midrule
    \multicolumn{7}{c}{Models}                                                                                                                                                                                                                          \\ \midrule
    Linear                      & {\color[HTML]{009901} \checkmark} & {\color[HTML]{009901} \checkmark} & {\color[HTML]{009901} \checkmark} & {\color[HTML]{CB0000} \xmark}     & {\color[HTML]{009901} \checkmark} & {\color[HTML]{009901} \checkmark} \\ 
    Logistic                    & {\color[HTML]{009901} \checkmark} & {\color[HTML]{009901} \checkmark} & {\color[HTML]{009901} \checkmark} & {\color[HTML]{CB0000} \xmark}     & {\color[HTML]{009901} \checkmark} & {\color[HTML]{009901} \checkmark} \\ 
    Quantile                    & {\color[HTML]{009901} \checkmark} & {\color[HTML]{CB0000} \xmark}     & {\color[HTML]{CB0000} \xmark}     & {\color[HTML]{009901} \checkmark} & {\color[HTML]{CB0000} \xmark}     & {\color[HTML]{009901} \checkmark} \\ \midrule
    \multicolumn{7}{c}{Penalizations}                                                                                                                                                                                                                   \\ \midrule
    Ridge                       & {\color[HTML]{009901} \checkmark} & {\color[HTML]{009901} \checkmark} & {\color[HTML]{009901} \checkmark} & {\color[HTML]{009901} \checkmark} & {\color[HTML]{009901} \checkmark} & {\color[HTML]{009901} \checkmark} \\ 
    Adaptive ridge              & {\color[HTML]{009901} \checkmark} & {\color[HTML]{CB0000} \xmark}     & {\color[HTML]{009901} \checkmark} & {\color[HTML]{CB0000} \xmark}     & {\color[HTML]{CB0000} \xmark}     & {\color[HTML]{CB0000} \xmark}     \\ 
    Lasso                       & {\color[HTML]{009901} \checkmark} & {\color[HTML]{009901} \checkmark} & {\color[HTML]{009901} \checkmark} & {\color[HTML]{009901} \checkmark} & {\color[HTML]{009901} \checkmark} & {\color[HTML]{009901} \checkmark} \\ 
    Adaptive lasso              & {\color[HTML]{009901} \checkmark} & {\color[HTML]{CB0000} \xmark}     & {\color[HTML]{009901} \checkmark} & {\color[HTML]{CB0000} \xmark}     & {\color[HTML]{CB0000} \xmark}     & {\color[HTML]{CB0000} \xmark}     \\ 
    Group lasso                 & {\color[HTML]{009901} \checkmark} & {\color[HTML]{009901} \checkmark} & {\color[HTML]{CB0000} \xmark}     & {\color[HTML]{009901} \checkmark} & {\color[HTML]{009901} \checkmark} & {\color[HTML]{CB0000} \xmark}     \\ 
    Adaptive group lasso        & {\color[HTML]{009901} \checkmark} & {\color[HTML]{CB0000} \xmark}     & {\color[HTML]{CB0000} \xmark}     & {\color[HTML]{CB0000} \xmark}     & {\color[HTML]{CB0000} \xmark}     & {\color[HTML]{CB0000} \xmark}     \\ 
    Sparse group lasso          & {\color[HTML]{009901} \checkmark} & {\color[HTML]{009901} \checkmark} & {\color[HTML]{CB0000} \xmark}     & {\color[HTML]{CB0000} \xmark}     & {\color[HTML]{009901} \checkmark} & {\color[HTML]{CB0000} \xmark}     \\ 
    Adaptive sparse group lasso & {\color[HTML]{009901} \checkmark} & {\color[HTML]{CB0000} \xmark}     & {\color[HTML]{CB0000} \xmark}     & {\color[HTML]{CB0000} \xmark}     & {\color[HTML]{CB0000} \xmark}     & {\color[HTML]{CB0000} \xmark}     \\
    Elastic net                 & {\color[HTML]{009901} \checkmark} & {\color[HTML]{009901} \checkmark} & {\color[HTML]{009901} \checkmark} & {\color[HTML]{009901} \checkmark} & {\color[HTML]{009901} \checkmark} & {\color[HTML]{009901} \checkmark} \\ 
    Adaptive elastic net        & {\color[HTML]{009901} \checkmark} & {\color[HTML]{CB0000} \xmark}     & {\color[HTML]{009901} \checkmark} & {\color[HTML]{CB0000} \xmark}     & {\color[HTML]{CB0000} \xmark}     & {\color[HTML]{CB0000} \xmark}     \\ \midrule
    \multicolumn{7}{c}{Adaptive weight estimation alternatives}                                                                                                                                                                                         \\ \midrule
    Available methodologies     & {\color[HTML]{009901} \checkmark} & {\color[HTML]{CB0000} \xmark}     & {\color[HTML]{CB0000} \xmark}     & {\color[HTML]{CB0000} \xmark}     & {\color[HTML]{CB0000} \xmark}     & {\color[HTML]{CB0000} \xmark}     \\ 
    \bottomrule
\end{tabular}
}
\end{table}



\section{Theoretical background}\label{sec:theory}

Let $\mathbb{D} = \{(y_i, \bm{x}_i)\}_{i=1}^n$ be a sample of $n$ observations, where $y_i \in \mathbb{R}$ is the response for the $i$-th observation (or $y_i \in \{0,1\}$ for binary classification) and $\bm{x}_i = (x_{i1}, \ldots, x_{ip})^T \in \mathbb{R}^p$ is the corresponding vector of $p$ covariates. The objective is to estimate the relationship between $\bm{x}_i$ and $y_i$. The \pkg{asgl} package considers models where the response or its transformation is related to a linear combination of the predictors, typically $\beta_0 + \bm{x}_i^T\bm{\beta}$, where $\beta_0$ is an optional intercept term and $\bm{\beta} \in \mathbb{R}^p$ is the vector of slope coefficients.

\subsection{Model formulations and loss functions}\label{sec:loss_functions}

The \pkg{asgl} package supports three primary types of regression models, which are distinguished by their respective loss functions $R(\beta_0, \bm{\beta})$. The model coefficients are typically estimated by minimizing this loss function:
$(\hat{\beta}_0, \hat{\bm{\beta}}) = \text{arg}\min_{\beta_0, \bm{\beta}} R(\beta_0, \bm{\beta})$.

\subsubsection{Linear Regression (Least Squares)}\label{sec:ls}
In linear regression, the model assumes the relationship $y_i = \beta_0 + \bm{x}_i^T\bm{\beta} + \varepsilon_i$, where $\varepsilon_i$ is an error term. The coefficients are estimated by minimizing the sum of squared errors (L2 loss):
\begin{equation}\label{eq:loss_lm}
R_{\text{LS}}(\beta_0, \bm{\beta}) = \frac{1}{n}\sum_{i=1}^n (y_i - (\beta_0 + \bm{x}_i^T\bm{\beta}))^2.
\end{equation}

\subsubsection{Logistic Regression}\label{sec:logit}
Logistic regression, a type of Generalized Linear Model (GLM) that models Bernoulli or binomial distributed data, is employed for binary classification problems where the response variable $y_i \in \{0, 1\}$. It models the conditional probability of $y_i=1$ as $P(y_i=1 | \bm{x}_i) = \sigma(\beta_0 + \bm{x}_i^T\bm{\beta})$, where $\sigma(z) = 1/(1+e^{-z})$ is the sigmoid function. The coefficients are estimated by minimizing the negative log-likelihood:
\begin{equation}\label{eq:loss_logistic}
R_{\text{Logit}}(\beta_0, \bm{\beta}) = -\frac{1}{n} \sum_{i=1}^n \left[ y_i \log(\sigma(\beta_0 + \bm{x}_i^T\bm{\beta})) + (1-y_i) \log(1-\sigma(\beta_0 + \bm{x}_i^T\bm{\beta})) \right].
\end{equation}

\subsubsection{Quantile Regression}\label{sec:qr}
Quantile regression, introduced by \citet{Koenker1978}, allows for the estimation of the conditional $\tau$-th quantile of the response variable, denoted as $Q_{y_i}(\tau | \bm{x}_i) = \beta_0 + \bm{x}_i^T\bm{\beta}$. This method is particularly advantageous due to its robustness to outliers and its ability to model different parts of the conditional distribution of $y_i$, making it suitable for data with heterogeneous variance. The coefficients are determined by minimizing the sum of asymmetrically weighted absolute errors:
\begin{equation}\label{eq:loss_qr}
R_{\text{QR}}^{(\tau)}(\beta_0, \bm{\beta}) = \frac{1}{n}\sum_{i=1}^n \rho_{\tau}(y_i - (\beta_0 + \bm{x}_i^T\bm{\beta})),
\end{equation}
where $\rho_{\tau}(u) = u(\tau - I(u<0))$ is the pinball loss, $I(\cdot)$ is the indicator function, and $\tau \in (0,1)$ is the specific quantile of interest.

\subsection{Penalized regression framework}\label{sec:penalized_regression}
In scenarios involving high-dimensional data (where $p > n$) or multicollinearity among predictors, standard estimation procedures can lead to unstable or non-unique solutions. Penalized regression methods address these issues by adding a penalty term $P(\bm{\beta})$ to the loss function. The general optimization problem for penalized regression is:
\begin{equation}\label{eq:general_penalized}
(\hat{\beta}_0, \hat{\bm{\beta}}) = \text{argmin}_{\beta_0 \in \mathbb{R}, \bm{\beta} \in \mathbb{R}^p} \left\{ R(\beta_0, \bm{\beta}) + P(\bm{\beta}) \right\},
\end{equation}
where $R(\beta_0, \bm{\beta})$ is one of the loss functions defined in Section~\ref{sec:loss_functions}. Consistent with standard practice, the intercept term $\beta_0$, if included in the model (i.e., if \code{fit_intercept=True}), is not subjected to penalization. The \pkg{asgl} package implements a variety of penalization techniques, which are summarized in Table~\ref{tab:penalties}. The main hyperparameters controlling these penalties in \pkg{asgl} are \code{lambda1} ($\lambda$, overall penalty strength) and \code{alpha} ($\alpha$, mixing proportion, its role depends on the specific penalization).

\begin{table}[htb]
  \centering
  \caption{Penalization terms $P(\bm{\beta})$ implemented in \pkg{asgl}. Here, $\bm{\beta} \in \mathbb{R}^p$ are the slope coefficients. For group penalties, $K$ is the number of groups, $p_l$ is the size of group $l$, and $\bm{\beta}^l$ are the coefficients in group $l$. The hyperparameters $\lambda$ and $\alpha$ correspond to parameters in the \code{Regressor} class. Weights $\tilde{w}_j$ and $\tilde{v}_l$ are used in adaptive versions.}
  \label{tab:penalties}
  \resizebox{\textwidth}{!}{
  \begin{tabular}{lll}
    \toprule
    Penalization & $P(\bm{\beta})$ & hyperparameters \\
    \midrule
    \multicolumn{3}{l}{Standard Penalizations} \\ \midrule
    Lasso & $\lambda\|\bm{\beta}\|_1$ & $\lambda$ \\
    Ridge & $\lambda\|\bm{\beta}\|_2^2$ & $\lambda$ \\
    Group Lasso (GL) & $\lambda\sum_{l=1}^K \sqrt{p_l} \|\bm{\beta}^l\|_2$ & $\lambda$, \code{group\_index} \\
    Sparse Group Lasso (SGL) & $\lambda\left( \alpha \|\bm{\beta}\|_1 + (1-\alpha) \sum_{l=1}^K \sqrt{p_l} \|\bm{\beta}^l\|_2 \right)$ & $\lambda$, $\alpha$, \code{group\_index} \\
    \midrule
    \multicolumn{3}{l}{Adaptive Penalizations (weights $\tilde{w}_j, \tilde{v}_l$ from Section~\ref{sec:adaptive_weights})} \\
    \midrule
    Adaptive Lasso (ALasso) & $\lambda\sum_{j=1}^p \tilde{w}_j |\beta_j|$ & $\lambda$, $\tilde{w}_j$ \\
    Adaptive Ridge (ARidge) & $\lambda\sum_{j=1}^p \tilde{w}_j \beta_j^2$ & $\lambda$, $\tilde{w}_j$ \\
    Adaptive Group Lasso (AGL) & $\lambda\sum_{l=1}^K \tilde{v}_l \sqrt{p_l} \|\bm{\beta}^l\|_2$ & $\lambda$, \code{group\_index}, $\tilde{w}_j$ \\
    Adaptive SGL (ASGL)& $\lambda\left( \alpha \sum_{j=1}^p \tilde{w}_j |\beta_j| + (1-\alpha) \sum_{l=1}^K \tilde{v}_l \sqrt{p_l} \|\bm{\beta}^l\|_2 \right)$ & $\lambda$, $\alpha$, \code{group\_index}, $\tilde{w}_j$ \\
    \bottomrule
  \end{tabular}%
  }
\end{table}

\subsubsection{Lasso}
The Lasso (Least Absolute Shrinkage and Selection Operator), proposed by \citet{Tibshirani1996}, employs an $\ell_1$ penalty on the coefficients (see Table~\ref{tab:penalties}). This penalty has the property of shrinking some coefficients to exactly zero, thereby performing variable selection simultaneously with estimation.

\subsubsection{Ridge Regression}
Ridge regression, introduced by \citet{hoerl_ridge_1970}, utilizes an $\ell_2$-squared penalty (Table~\ref{tab:penalties}). This form of regularization shrinks the coefficients towards zero, which is particularly effective for stabilizing estimates when predictors are highly correlated (multicollinearity). However, Ridge regression does not typically set coefficients to exactly zero, so it does not perform variable selection.

\subsubsection{Group Lasso}
In situations where predictors possess a natural grouping structure (e.g., dummy variables from a categorical predictor, or genes in a pathway), the Group Lasso \citep{Yuan2006} can be applied. This penalty (Table~\ref{tab:penalties}) encourages sparsity at the group level, meaning that entire groups of variables are either included in or excluded from the model. The penalty involves summing the $\ell_2$ norms of the coefficient vectors within each group, typically weighted by the square root of the group size $\sqrt{p_l}$.

\subsubsection{Sparse Group Lasso}
The Sparse Group Lasso (SGL), described by \citet{friedman_note_2010} extends the Group Lasso by incorporating an additional $\ell_1$ penalty on individual coefficients within the groups (Table~\ref{tab:penalties}). This allows for sparsity both at the group level (some groups entirely excluded) and at the individual coefficient level within the selected groups (some variables within an active group can have zero coefficients). The \code{alpha} parameter in \pkg{asgl} (when \code{penalization="sgl"}) controls the mixture $\alpha$ between the individual $\ell_1$ sparsity and the group $\ell_2$ sparsity.

\subsubsection{Adaptive penalizations}
\begin{figure*}
	\centering
	\includegraphics[width=12cm]{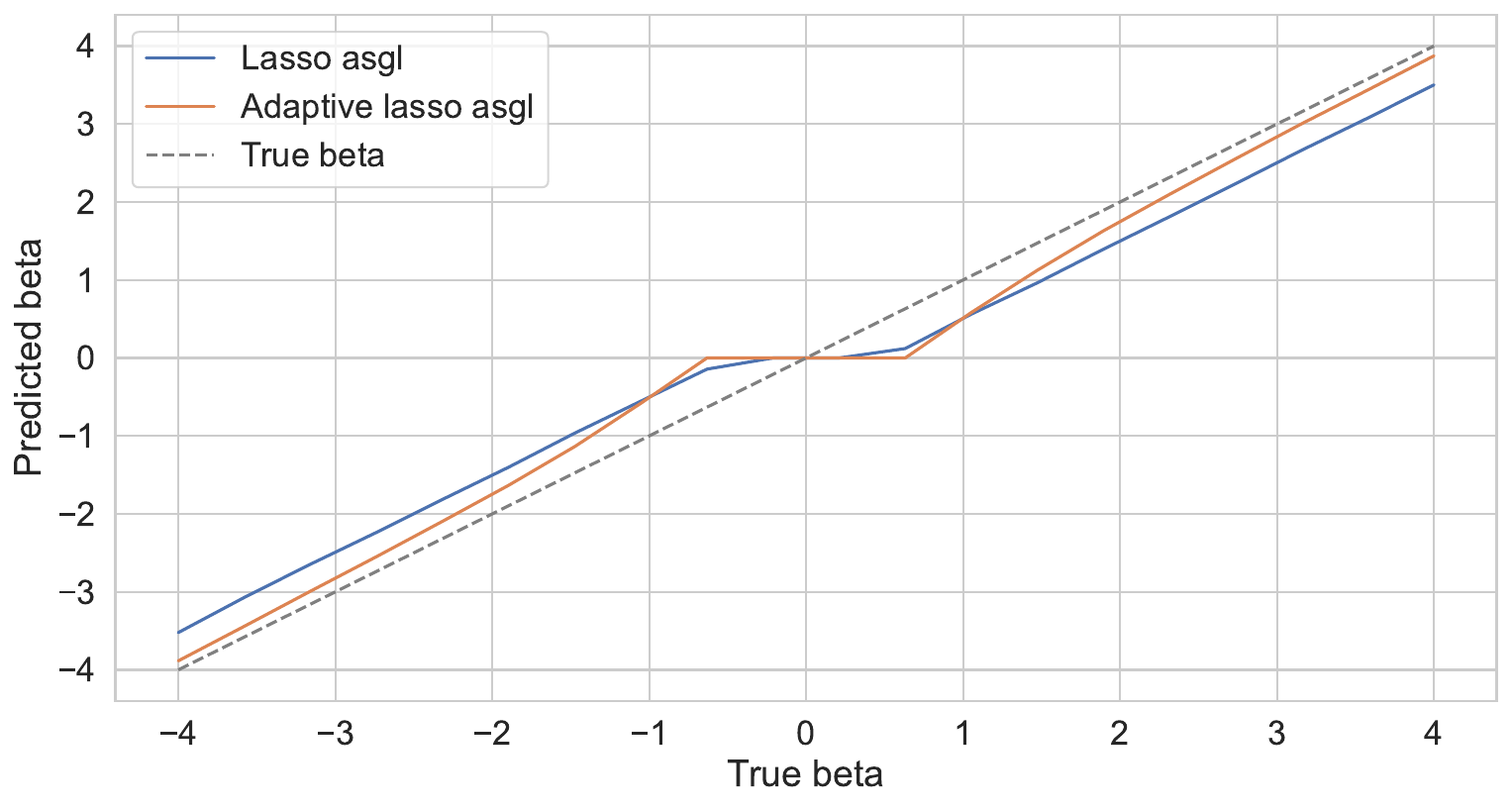}
	\caption{Thresholding functions for lasso and adaptive lasso, illustrating the adaptive method's bias reduction. By using variable-specific weights, adaptive lasso applies less shrinkage to coefficients with larger estimated magnitudes (its function is closer to the identity line for larger inputs), thereby reducing estimation bias for important variables while still promoting sparsity.}
	\label{fig:thresholding}
\end{figure*}

Standard penalization methods apply a uniform penalty constraint across all coefficients (or groups). This can lead to biased estimates for coefficients with large true magnitudes and may not always yield optimal variable selection. \citet{Zou2006a} introduced the adaptive Lasso, which assigns different penalty strengths to different coefficients by incorporating variable-specific weights into the $\ell_1$ penalty. The intuition is to penalize coefficients with small initial estimates more heavily, and those with large initial estimates less so. This approach often leads to improved statistical properties, such as consistency in variable selection (achieving the oracle property under certain conditions) and reduced estimation bias for the non-zero coefficients. Figure~\ref{fig:thresholding} illustrates how adaptive Lasso can mitigate the bias of standard Lasso.
This adaptive concept can be generalized to other types of penalizations. The adaptive versions of the penalties available in \pkg{asgl} are detailed in Table~\ref{tab:penalties}. These formulations incorporate pre-computed weights $\tilde{w}_j$ for individual coefficients and/or $\tilde{v}_l$ for groups of coefficients. The critical step of estimating these weights is discussed in Section~\ref{sec:adaptive_weights}. The development of adaptive sparse group lasso (ASGL) and methods for its weight calculation were detailed in~\cite{Mendez-Civieta2020}.

\subsection{Adaptive weight estimation}\label{sec:adaptive_weights}
The superior performance of adaptive penalizations is critically dependent on the appropriate choice of adaptive weights, $\tilde{w}_j$ for individual coefficients and $\tilde{v}_l$ for coefficient groups. These weights are typically defined as inversely proportional to the magnitude of initial coefficient estimates:
\begin{equation}\label{eq:weight_formula}
\tilde{w}_j = \frac{1}{|\hat{\beta}_j^{\text{init}}|^{\gamma_1}} \quad \text{and} \quad \tilde{v}_l = \frac{1}{\|\hat{\bm{\beta}}_l^{\text{init}}\|_2^{\gamma_2}},
\end{equation}
where $\hat{\beta}_j^{\text{init}}$ is an initial estimate for the $j$-th coefficient, $\hat{\bm{\beta}}_l^{\text{init}}$ is the vector of initial coefficient estimates for group $l$, and $\gamma_1, \gamma_2 > 0$ are positive constants (commonly 1 or 2) that control the adaptiveness. A small tolerance \code{weight_tol} is added to the denominator to prevent division by zero.

A significant practical challenge, especially in high-dimensional settings ($p > n$), is obtaining reliable initial estimates $\hat{\bm{\beta}}^{\text{init}}$. Standard unpenalized estimation (e.g., Ordinary Least Squares or Maximum Likelihood Estimation) is often infeasible or yields highly variable estimates when $p > n$. While some software packages offer adaptive penalizations (e.g., \pkg{glmnet} in \proglang{R} for adaptive Lasso), they require the user to supply these weights, thereby not fully addressing this practical hurdle.

A distinctive feature and core contribution of the \pkg{asgl} package is its implementation of several built-in methodologies for estimating these adaptive weights automatically. This makes adaptive penalizations practically accessible even in high-dimensional contexts. The methods, some of which were introduced or extended by~\citet{Mendez-Civieta2020} in the context of adaptive sparse group lasso, are available for all adaptive penalizations in \pkg{asgl} and include:

\begin{itemize}
    \item \textbf{Unpenalized} (\code{weight_technique="unpenalized"}): Derives weights from coefficients of an unpenalized model (using one of the loss functions in Section~\ref{sec:loss_functions}). This method is generally suitable only for low-dimensional data ($p < n$).
    \item \textbf{Ridge-based} (\code{weight_technique="ridge"}): Employs coefficients from an initial Ridge regression fit (Section~\ref{sec:penalized_regression}) to calculate weights. Ridge regression provides stable, non-sparse initial estimates that are well-suited for weight calculation, even in the presence of multicollinearity or when $p > n$.
    \item \textbf{Lasso-based} (\code{weight_technique="lasso"}): Uses coefficients from an initial Lasso fit (Section~\ref{sec:penalized_regression}). This approach can yield sparse initial estimates, thereby focusing the adaptive weights on variables selected by the preliminary Lasso model.
    \item \textbf{PCA-based} (\code{weight_technique="pca_pct"} or \code{"pca_1"}): For high-dimensional data, Principal Component Analysis (PCA) of the covariate matrix $\bm{X}$ can be used to obtain robust initial estimates for weight calculation.
        \begin{itemize}
            \item \code{"pca_pct"}: An unpenalized regression is performed on a subset of principal component scores. These scores correspond to components that collectively explain a user-specified percentage (\code{variability_pct}) of the total variance in $\bm{X}$. The resulting regression coefficients are then transformed back to the original feature space to yield $\hat{\bm{\beta}}^{\text{init}}$.
            \item \code{"pca_1"}: The loadings of the first principal component of $\bm{X}$ are used directly to form (or derive) the initial coefficient estimates $\hat{\bm{\beta}}^{\text{init}}$.
        \end{itemize}
        These PCA-based techniques facilitate stable weight estimation in $p > n$ scenarios by first reducing dimensionality in an unsupervised manner.
    \item \textbf{PLS-based} (\code{weight_technique="pls_pct"} or \code{"pls_1"}): Analogous to PCA-based methods, Partial Least Squares (PLS) regression is utilized to derive initial weights. PLS components are constructed to maximize the covariance between linear combinations of predictors and the response variable $y$, potentially making them more relevant for predictive modeling than PCA components.
        \begin{itemize}
            \item \code{"pls_pct"}: Uses an unpenalized regression on PLS component scores that explain a specified \code{variability_pct} of the variance of the predictors
            \item \code{"pls_1"}: Employs loadings from the first PLS component.
        \end{itemize}
    \item \textbf{Sparse PCA-based} (\code{weight_technique="sparse_pca"}): This method uses Sparse PCA, as proposed by \citet{Zou2006}, to obtain sparse principal component loadings. These sparse loadings are then used to derive $\hat{\bm{\beta}}^{\text{init}}$, similar to the standard PCA-based methods but with an inherent emphasis on identifying sparse combinations of the original features for weight estimation.
\end{itemize}
The provision of these diverse and automated adaptive weight calculation techniques significantly enhances the practical applicability and power of adaptive penalization methods, especially for the complex, high-dimensional datasets frequently encountered in modern research. This comprehensive suite of options for weight estimation is a key strength of the \pkg{asgl} package, and already showed its benefits in~\cite{Mendez-Civieta2020}.


\section[Python implementation]{\proglang{Python} implementation}\label{sec:python_implementation}

\pkg{asgl} is an open-source \proglang{python} package designed around a central class named \class{Regressor}, which serves as the main interface for users to fit the various regression models and penalizations described in Section \ref{sec:theory}. This class adheres to the estimator guidelines provided by \pkg{scikit-learn} \citep{scikit-learn}, the leading library for machine learning in \proglang{Python}. By following these guidelines, the \class{Regressor} class ensures seamless integration with the \pkg{scikit-learn} ecosystem. This grants users access to a wide range of familiar tools for model evaluation, hyperparameter tuning, cross-validation (e.g., via \class{GridSearchCV} or \class{RandomizedSearchCV}), prediction, and error computation, allowing \pkg{asgl} models to be easily incorporated into existing machine learning pipelines.

\subsection[The Regressor class]{The \code{Regressor} class}

The \code{Regressor} class in the \pkg{asgl} package is versatile and supports a range of regression models and penalizations, making it suitable for various statistical modeling needs. An instance of the class is created by calling the constructor with specific parameters to define the desired model configuration. The default parameters for the \class{Regressor} class are:

\begin{CodeChunk}
\begin{CodeInput}
model = asgl.Regressor(
    model="lm", penalization="lasso", quantile=0.5, 
    fit_intercept=True, lambda1=0.1, alpha=0.5, solver="CLARABEL", 
    weight_technique="pca_pct", individual_power_weight=1, 
    group_power_weight=1, variability_pct=0.9, lambda1_weights=0.1, 
    spca_alpha=1e-5, spca_ridge_alpha=1e-2, 
    individual_weights=None, group_weights=None, 
    weight_tol=1e-4, tol=1e-3)
\end{CodeInput}
\end{CodeChunk} 

Below are the parameters and their descriptions:

\begin{itemize}
  \item \textbf{model:} \code{str}, \code{default="lm"} \\
  Specifies the type of model to fit. Options are \code{"lm"} for linear regression (Section \ref{sec:ls}), \code{"qr"} for quantile regression (Section \ref{sec:qr}), and \code{"logit"} for logistic regression for binary classification (Section \ref{sec:logit}).
  
  \item \textbf{penalization:} \code{str} or \code{None}, \code{default="lasso"} \\
  Defines the type of penalization to apply. Options include \code{"lasso"}, \code{"ridge"}, \code{"gl"} (group lasso), \code{"sgl"} (sparse group lasso), and their adaptive counterparts: \code{"alasso"} (adaptive lasso), \code{"aridge"} (adaptive ridge), \code{"agl"} (adaptive group lasso), \code{"asgl"} (adaptive sparse group lasso). If \code{None}, an unpenalized model is fitted. See Table \ref{tab:penalties} for formulations.
  
  \item \textbf{quantile:} \code{float}, \code{default=0.5} \\
  The quantile level $\tau \in (0,1)$ for quantile regression models. This parameter is only active if \code{model="qr"}.
  
  \item \textbf{fit\_intercept:} \code{bool}, \code{default=True} \\
  Whether to calculate an intercept term ($\beta_0$) for the model. If \code{True}, the intercept is fitted; otherwise, it is assumed to be zero. The intercept is not penalized.
  
  \item \textbf{lambda1:} \code{float}, \code{default=0.1} \\ 
  The main regularization parameter that multiplies the overall penalty term $P(\bm{\beta})$ (see Equation \ref{eq:general_penalized} and Table \ref{tab:penalties}). It must be a non-negative float, i.e., in $[0,\infty)$. Larger values correspond to stronger penalization.
 
  \item \textbf{alpha:} \code{float}, \code{default=0.5} \\
  A mixing parameter used in Sparse Group Lasso and Adaptive Sparse Group Lasso penalizations (see Table \ref{tab:penalties}). It balances individual $\ell_1$ sparsity and group $\ell_2$ sparsity, where \code{alpha=1} enforces individual sparsity (like Lasso) and \code{alpha=0} enforces group sparsity (like Group Lasso). Must be in $[0,1]$.
  
  \item \textbf{solver:} \code{str}, \code{default="default"} \\
  Specifies the solver to be used by the underlying \pkg{cvxpy} optimization backend. The \code{"CLARABEL"} setting uses \pkg{cvxpy}'s open source convex optimization solver \code{CLARABEL}. Users can list the available solvers with \code{cvxpy.installed_solvers()} and specify one if needed. \pkg{cvxpy} includes open source and commercial solvers.
  
  \item \textbf{weight\_technique:} \code{str}, \code{default="pca\_pct"} \\
  The technique for estimating adaptive weights if an adaptive penalization is chosen and custom weights are not provided. Options are detailed in Section \ref{sec:adaptive_weights} and include: \code{"pca_1"}, \code{"pca_pct"}, \code{"pls_1"}, \code{"pls_pct"}, \code{"lasso"}, \code{"ridge"}, \code{"unpenalized"}, and \code{"sparse_pca"}. For low-dimensional problems ($p < n$), \code{"unpenalized"} or \code{"ridge"} are often suitable. For high-dimensional problems ($p > n$), \code{"pca_pct"} (default)is generally recommended.
  
  \item \textbf{individual\_power\_weight:} \code{float}, \code{default=1} \\
 The exponent $\gamma_1$ applied to the inverse of initial coefficient estimates when calculating individual adaptive weights $\tilde{w}_j$ (see Equation \ref{eq:weight_formula}). Active for adaptive penalizations involving individual coefficient weights (e.g., \code{"alasso"}, \code{"aridge"}, \code{"asgl"}).
  
  \item \textbf{group\_power\_weight:} \code{float}, \code{default=1} \\
  The exponent $\gamma_2$ applied to the inverse of initial group norm estimates when calculating group adaptive weights $\tilde{v}_l$ (see Equation \ref{eq:weight_formula}). Active for adaptive penalizations involving group weights (e.g., \code{"agl"}, \code{"asgl"})
  
  \item \textbf{variability\_pct:} \code{float}, \code{default=0.9} \\
  The percentage of variance to be explained by the selected components when \\ \code{weight_technique} is \code{"pca_pct"}, \code{"pls_pct"}, or \code{"sparse_pca"}. Must be in $(0,1]$.
  
  \item \textbf{lambda1\_weights:} \code{float}, \code{default=0.1} \\
  The regularization parameter for the initial Lasso or Ridge model used in adaptive weight estimation, applicable when \code{weight_technique="lasso"} or \\ \code{weight_technique="ridge"} respectively.
  
  \item \textbf{spca\_alpha:} \code{float}, \code{default=1e-5} \\
  The L1 regularization parameter for Sparse PCA if \code{weight_technique="sparse_pca"}. Corresponds to the \code{alpha} parameter in \pkg{scikit-learn}'s \class{SparsePCA}.
  
  \item \textbf{spca\_ridge\_alpha:} \code{float, default=1e-2} \\
  The Ridge regularization parameter for Sparse PCA if \\ \code{weight_technique="sparse_pca"}. Corresponds to the \code{ridge_alpha} parameter in \pkg{scikit-learn}'s \class{SparsePCA}.
  
  \item \textbf{individual\_weights:} \code{array} or \code{None}, \code{default=None} \\
  User-provided custom individual weights $\tilde{w}_j$ for adaptive penalizations. If supplied, this array overrides the internal weight estimation process specified by \code{weight_technique}. Must be a 1D array of non-negative floats with length equal to the number of features $p$.
  
  \item \textbf{group\_weights:} \code{array} or \code{None}, \code{default=None} \\
  User-provided custom group weights $\tilde{v}_l$ for adaptive penalizations with group structures. If supplied, this array overrides internal group weight estimation. Must be a 1D array of non-negative floats with length equal to the number of groups $K$ (as defined by \code{group_index}).
  
  \item \textbf{tol:} \code{float}, \code{default=1e-3} \\
  Tolerance for treating estimated coefficients as zero. Coefficients with absolute values smaller than \code{tol} are considered zero.
  
  \item \textbf{weight\_tol:} \code{float}, \code{default=1e-4} \\
  A small tolerance value added to the denominator during adaptive weight calculation (Equation \ref{eq:weight_formula}) to prevent division by zero if an initial coefficient estimate is exactly zero.
\end{itemize}

The main methods included in the \code{Regressor} class object are \code{fit} and \code{predict}.

\subsubsection[fit function]{\code{fit} function}
The \code{fit} method is used to train the model specified by the \class{Regressor} instance.
\begin{CodeChunk}
\begin{CodeInput}
model.fit(X, y, group_index=None)
\end{CodeInput}
\end{CodeChunk}
Its parameters are:
\begin{itemize}
    \item \code{X}: A 2D \code{numpy.ndarray} of shape \code{(n_samples, n_features)} representing the predictor variables.
    \item \code{y}: A 1D \code{numpy.ndarray} of shape \code{(n_samples,)} representing the response variable.
    \item \code{group_index}: A 1D \code{numpy.ndarray} of shape \code{(n_features,)}, default=\code{None}. This array specifies the group membership for each feature. For example, \\ \code{group_index=numpy.array([1,1,2,2,3])} indicates that the first two features belong to group 1, the next two to group 2, and the last feature to group 3. This parameter is required only when using group-based penalizations (\code{"gl"}, \code{"sgl"}, \code{"agl"}, \code{"asgl"}). Groups should be labeled with consecutive positive integers.
\end{itemize}

After the \code{fit} method is successfully called, the \class{Regressor} generates attributes:

\begin{itemize}
    \item \textbf{coef\_}: A 1D \code{numpy.ndarray} of dimension \code{n_features} storing the slope coefficients $\hat{\bm{\beta}}$.
    \item \textbf{intercept\_}: A \code{numpy.float} number. If \code{fit_intercept=True}, this is the intercept $\hat{\beta}_0$. If \code{fit_intercept=False} this takes value 0.
    \item \textbf{is\_fitted\_}: A boolean indicating if the model has been fitted.
    \item \textbf{n\_features\_in\_}: A \code{numpy.integer} number indicating the number of columns of \code{X} used in training.
\end{itemize}

\subsubsection[predict method]{\code{predict} method}
The \code{predict} method is used to make predictions on new data using the fitted model.
\begin{CodeChunk}
\begin{CodeInput}
predictions = model.predict(X)
\end{CodeInput}
\end{CodeChunk}
Its parameter is:
\begin{itemize}
    \item \code{X}: A 2D \code{numpy.ndarray} of shape \code{(n_new_samples, n_features)} for which predictions are desired. The number of features must match that of the training data.
\end{itemize}
The method returns a 1D \code{numpy.ndarray} of shape \code{(n_new_samples,)} containing the predicted values. For linear and quantile regression (\code{model="lm"} or \code{model="qr"})), this array contains the predicted response values. For logistic (\code{model="logit"}), this method returns the predicted class.

\subsubsection[decision function method for Logistic Regression]{\code{decision function} method for Logistic Regression}
This method is available only when \code{model="logit"}. It provides the raw output of the linear part of the model, i.e., $z = \beta_0 + \bm{x}^T\bm{\beta}$, before this value is passed through the sigmoid function. These values can be interpreted as the signed distance to the decision boundary.
\begin{CodeChunk}
\begin{CodeInput}
decision_scores = model.decision_function(X)
\end{CodeInput}
\end{CodeChunk}
Its parameter is:
\begin{itemize}
    \item \code{X}: A 2D \code{numpy.ndarray} of shape \code{(n_new_samples, n_features)} for which decision scores are desired.
\end{itemize}
The method returns a 1D \code{numpy.ndarray} of shape \code{(n_new_samples,)} containing the decision scores.

\subsubsection[predict proba method for Logistic Regression]{\code{predict proba} method for Logistic Regression}
This method is available only when \code{model="logit"}. It returns the probability estimates for each class.
\begin{CodeChunk}
\begin{CodeInput}
probabilities = model.predict_proba(X)
\end{CodeInput}
\end{CodeChunk}
Its parameter is:
\begin{itemize}
    \item \code{X}: A 2D \code{numpy.ndarray} of shape \code{(n_new_samples, n_features)} for which class probabilities are desired.
\end{itemize}
The method returns a 2D \code{numpy.ndarray} of shape \code{(n_new_samples, 2)}. Each row corresponds to a sample, and the columns contain the probability of the sample belonging to class 0 and class 1, respectively. The sum of probabilities in each row is 1.


\section{Usage example}\label{sec:examples}

This section provides a series of illustrations demonstrating the core functionalities of the \pkg{asgl} package, guiding the user from basic model fitting with standard penalizations to the more advanced adaptive methods that represent a key contribution of this work. The section is intended to show \pkg{asgl}'s capabilities, while methodological advancements can be seen in \cite{Mendez-Civieta2020}. \pkg{asgl} simplifies advanced methods like adaptive lasso by automating complex steps like high-dimensional weight estimation. This section underscores \pkg{asgl}'s user-friendly approach to adaptive penalized regression.

The \pkg{asgl} package is available from the \proglang{Python} Package Index (PyPI) at \url{https://pypi.org/project/asgl/} and the \proglang{GitHub} repository \url{https://github.com/alvaromc317/asgl/} and can be easily installed:

\begin{CodeChunk}
\begin{CodeInput}
pip install asgl
\end{CodeInput}
\end{CodeChunk}

The package relies on \pkg{cvxpy} \citep{Diamond2016} for solving the underlying convex optimization problems. Optional speed-ups may be available if commercial solvers compatible with \pkg{cvxpy} (e.g., \pkg{GUROBI}, \pkg{MOSEK}, \pkg{CPLEX}) are installed and detected at run-time; otherwise, \pkg{cvxpy} will transparently use suitable open-source solvers (e.g., \pkg{CLARABEL}, \pkg{OSQP}, \pkg{SCS}).

The following examples assume that necessary libraries, including \pkg{asgl} itself, \pkg{numpy}, and relevant modules from \pkg{scikit-learn} and \pkg{matplotlib}, have been installed.

\subsection{Example 1: standard penalizations and scikit-learn integration}

This first example serves to quickly demonstrate the basic syntax of \pkg{asgl}, its application to a standard model type —specifically linear regression with lasso penalization— and its seamless integration with the \pkg{scikit-learn} ecosystem for familiar tasks such as hyperparameter tuning. The example illustrates fundamental \code{fit} and \code{predict} usage and showcases hyperparameter optimization via \class{GridSearchCV}, highlighting that \pkg{asgl} estimators can be readily incorporated into existing \pkg{scikit-learn} workflows. This establishes a foundation before exploring the more advanced and unique features of the package in subsequent examples.

\begin{figure*}[htbp]
	\centering
	\includegraphics[width=12cm]{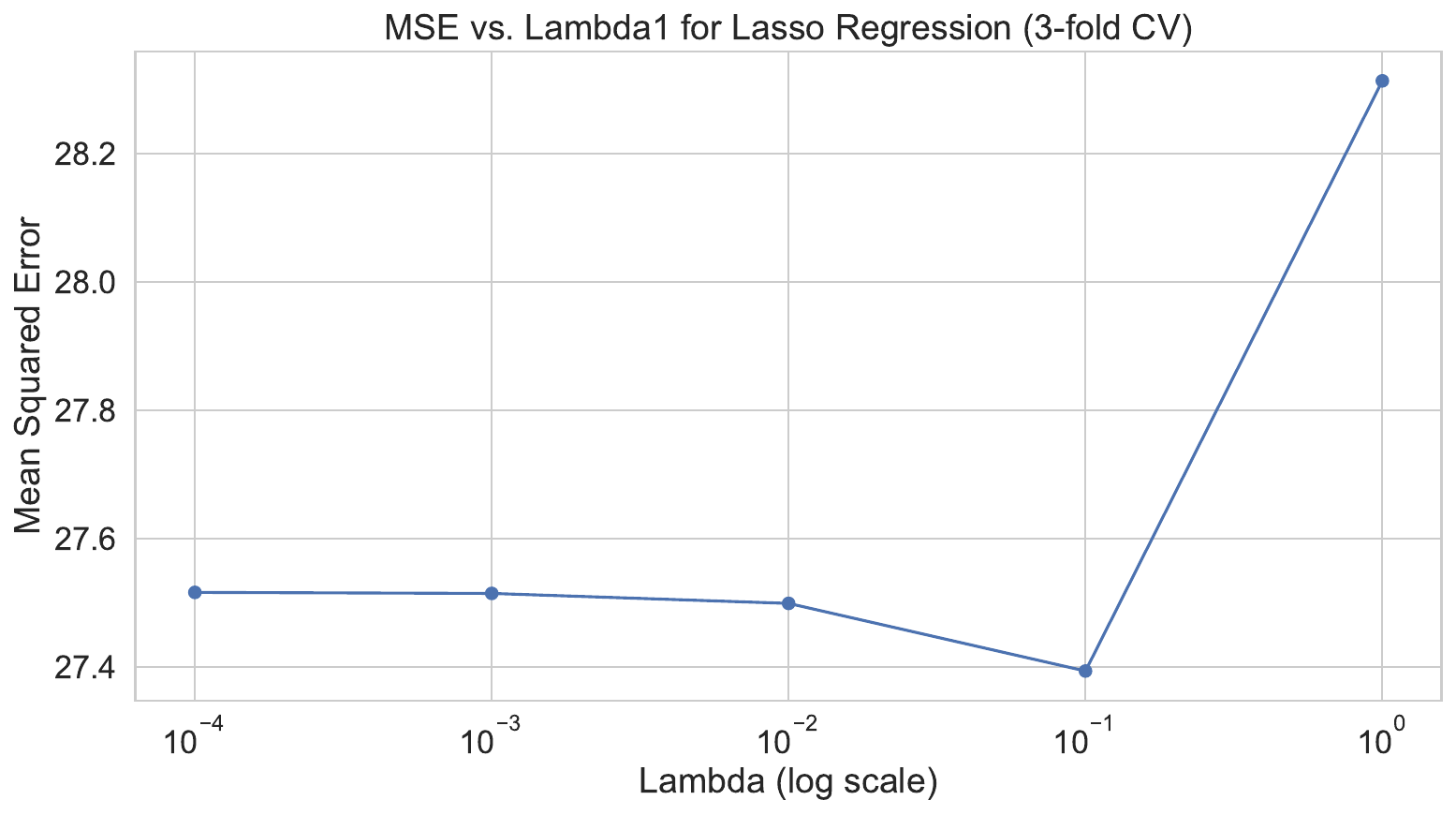}
	\caption{Example 1: MSE vs. \code{lambda1} for Lasso Regression (3-fold CV)}
	\label{fig:mse_lambda}
\end{figure*}

We begin by generating synthetic regression data using \code{make_regression} from \pkg{scikit-learn} and splitting it into training and testing sets. An \pkg{asgl} \class{Regressor} object is then instantiated for linear regression (\code{model="lm"}) with lasso penalization (\code{penalization="lasso"}). A parameter grid for the regularization strength \code{lambda1} is defined. \class{GridSearchCV} from \pkg{scikit-learn} is employed to find the optimal \code{lambda1} value using 3-fold cross-validation, with \code{scoring="neg_mean_squared_error"}. Figure~\ref{fig:mse_lambda} is generated to visualize the MSE across the different tested values of \code{lambda1}. Finally, predictions using the optimal \code{lambda1} are made on the test set, and the Mean Squared Error (MSE) is calculated.

\begin{CodeInput}
>>> import numpy as np
>>> from sklearn.datasets import make_regression
>>> from sklearn.model_selection import train_test_split, GridSearchCV
>>> from sklearn.metrics import mean_squared_error
>>> from asgl import Regressor
>>> import matplotlib.pyplot as plt

# Generate synthetic regression data
>>> X, y = make_regression(n_samples=1000, n_features=20, n_informative=5, 
...                        bias=10, noise=5, random_state=42)
>>> X_train, X_test, y_train, y_test = train_test_split(X, y, test_size=250, 
...                                                     random_state=42)

# Create a Regressor object for linear regression with Lasso penalization
>>> model = Regressor(model="lm", penalization="lasso")

# Define a parameter grid for hyper-parameter tuning
>>> lambda1_values =  [1e-4, 1e-3, 1e-2, 1e-1, 1]
>>> param_grid = {"lambda1": lambda1_values}

# Cross-validation process
>>> gscv = GridSearchCV(model, param_grid, scoring="neg_mean_squared_error", 
...                     n_jobs=-1, cv=3)
>>> gscv.fit(X_train, y_train)

# Plotting the cross-validation results
>>> plt.figure(figsize=(8, 3.5)) 
>>> plt.plot(lambda1_values, -1*gscv.cv_results_["mean_test_score"])
>>> plt.xscale("log")
>>> plt.xlabel("Lambda (log scale)")
>>> plt.ylabel("Mean Squared Error")
>>> plt.title("MSE vs. lambda1 for Lasso Regression (3-fold CV)")

# Make predictions using the best estimator from GridSearchCV
>>> predictions = gscv.predict(X_test)

# Evaluate the model's performance using mean squared error
>>> mse = np.round(mean_squared_error(predictions, y_test), 3)
>>> print(f"Mean Squared Error: {mse}")
\end{CodeInput}
\begin{CodeOutput}
Mean Squared Error: 25.03
\end{CodeOutput}

This example effectively demonstrates that \pkg{asgl} integrates smoothly into standard \pkg{scikit-learn} workflows. Similar approaches can be readily applied for other models available in \pkg{asgl}, such as logistic regression (\code{model="logit"}) or quantile regression (\code{model="qr"}), and other penalizations such as ridge (\code{penalization="ridge"}) or sparse group lasso (\code{penalization="sgl"}). The subsequent examples will delve into the package's more advanced and unique capabilities, particularly its comprehensive support for adaptive penalizations and the various built-in techniques for estimating the crucial adaptive weights, which form a core contribution of \pkg{asgl}.

\subsection{Example 2: Lasso vs. Adaptive Lasso with Increasing Dimensionality}

\begin{figure*}[htbp]
	\centering
	\includegraphics[width=\textwidth]{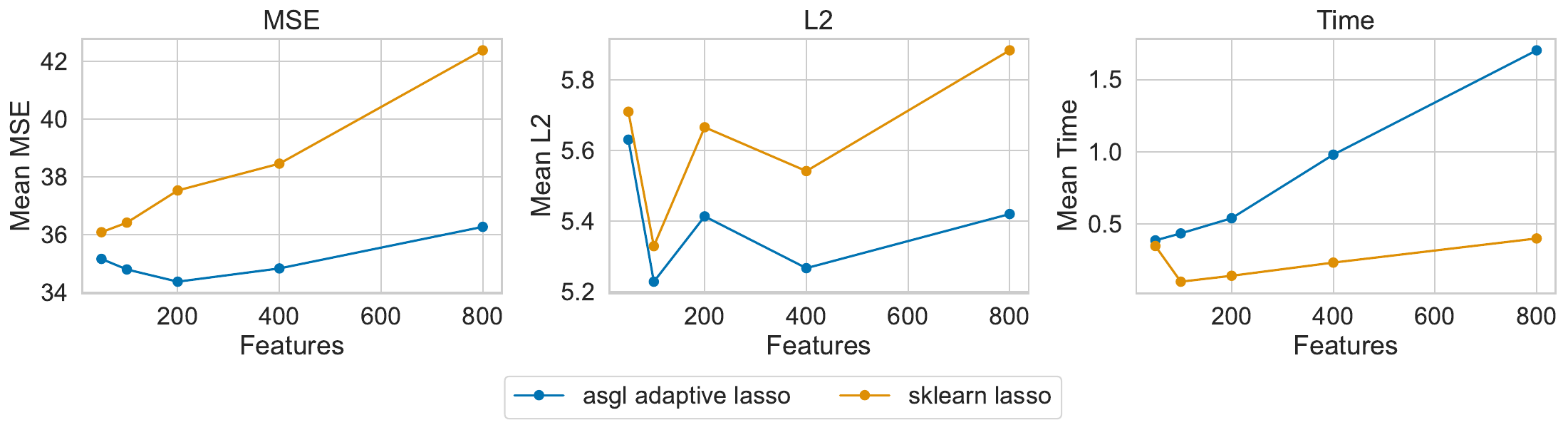}
	\caption{Example 2: Performance comparison between \pkg{scikit-learn}'s lasso and \pkg{asgl}'s adaptive lasso as the number of features increases. The number of training samples is fixed at 100. Metrics shown are Mean Squared Error (MSE), the $\ell_2$ norm of the coefficient error ($\|\bm{\beta}_{\text{true}}-\hat{\bm{\beta}}\|_2$), and execution time (in seconds). Results are averaged over 25 independent iterations.}
	\label{fig:lasso_vs_alasso}
\end{figure*}

This example is designed to showcase the advantages of adaptive lasso over the standard lasso, particularly in high-dimensional scenarios with collinear predictors. We simulate data where the number of features $p$ grows while the number of samples remains fixed, a setting that poses significant challenges for variable selection and estimation. The experiment compares the performance of \pkg{asgl}'s adaptive lasso against \pkg{scikit-learn}'s standard lasso implementation. The data generation process creates a collinear structure among predictors using the function \code{make_colinear_data} from the appendix, with a fixed number of observations ($n_{\text{train}}=100, n_{\text{test}}=100$) and a constant number of truly significant variables (25). We systematically increase the total number of features ($p \in \{50, 100, 200, 400, 800\}$) to observe how each model's performance evolves. The code below details the full simulation process, which iterates through each feature dimension setting 25 times to gather robust metrics. The aggregated results are then used to generate Figure~\ref{fig:lasso_vs_alasso}.

\begin{CodeInput}
>>> import numpy as np
>>> import pandas as pd
>>> import time
>>> from asgl import Regressor
>>> from sklearn.linear_model import Lasso
>>> from sklearn.model_selection import GridSearchCV, train_test_split
>>> from sklearn.metrics import mean_squared_error
>>> import matplotlib.pyplot as plt

# Simulation parameters
>>> n_repetitions = 25
>>> n_groups_arr = np.array([2, 4, 8, 16, 32]) # p = n_groups * 25
>>> param_grid = {'alpha': 10 ** np.arange(-1, 1, 0.05)}
>>> param_grid_alasso = {'lambda1': 10 ** np.arange(-1, 1, 0.05)} 

>>> results = []
>>> idx_seed = 0
>>> for n_groups in n_groups_arr:
...     n_features = n_groups * 25
...     print(f"Running simulations for p = {n_features}...")
...     for it in range(n_repetitions):
...         idx_seed += 1
...         X, y, beta_true = make_colinear_data(
...             n_obs=200, group_size=25, n_groups=n_groups, 
...             corr=0.7, seed=idx_seed)
...         X_train, X_test, y_train, y_test = train_test_split(
...             X, y, test_size=100, random_state=it)

...         # 1. Lasso using scikit-learn
...         start = time.perf_counter()
...         gscv_lasso_sk = GridSearchCV(
...             Lasso(), param_grid, scoring='neg_mean_squared_error', 
...             cv=3, n_jobs=-1)
...         gscv_lasso_sk.fit(X_train, y_train)
...         lasso_sk_pred = gscv_lasso_sk.predict(X_test)
...         results.append({
...             'Features': n_features, 'Model': 'sklearn lasso',
...             'MSE': mean_squared_error(lasso_sk_pred, y_test), 
...             'L2 betas': np.linalg.norm(beta_true -
...                             gscv_lasso_sk.best_estimator_.coef_),
...             'Exec time': time.perf_counter() - start})

...         # 2. Adaptive Lasso using asgl
...         start = time.perf_counter()
...         gscv_alasso_asgl = GridSearchCV(
...             Regressor(penalization='alasso'), param_grid_alasso, 
...             scoring='neg_mean_squared_error', cv=3, n_jobs=-1)
...         gscv_alasso_asgl.fit(X_train, y_train)
...         alasso_asgl_pred = gscv_alasso_asgl.predict(X_test)
...         results.append({
...             'Features': n_features, 'Model': 'asgl adaptive lasso',
...             'MSE': mean_squared_error(alasso_asgl_pred, y_test), 
...             'L2 betas': np.linalg.norm(beta_true - 
...                             gscv_alasso_asgl.best_estimator_.coef_),
...             'Exec time': time.perf_counter() - start})

# Results are then aggregated and plotted to generate Figure
>>> df_results = pd.DataFrame(results)
>>> print(df_results.groupby(['Features', 'Model']).mean())
\end{CodeInput}

The results of the simulation, averaged over 25 iterations, are presented in Figure~\ref{fig:lasso_vs_alasso}. The plots clearly illustrate the benefit of the adaptive penalization as the problem's dimensionality increases. For a low number of features ($p=50$ or $p=100$), both standard lasso and adaptive lasso perform comparably in terms of Mean Squared Error (MSE) and the $\ell_2$ norm of the coefficient error. However, as the number of features grows to 400 and 800, a distinct performance gap emerges. The standard lasso's MSE and coefficient error increase significantly, indicating a reduced ability to identify the true model and accurately estimate coefficients in the presence of many irrelevant, correlated predictors. In contrast, \pkg{asgl}'s adaptive lasso maintains a consistently low MSE and coefficient error, demonstrating its robustness and superior variable selection capabilities in challenging high-dimensional settings. While the adaptive lasso has a higher computational cost as shown in the execution time plot, these (1.5 seconds for the largest scenario with 800 variables) are very reasonable and this trade-off is often justified by the substantial gains in model accuracy and stability. This example highlights a key strength of \pkg{asgl}: providing an accessible implementation of adaptive methods that yield more reliable models in high-dimensional contexts.

\subsection{Example 3: lasso vs adaptive sparse group lasso in quantile regression}

This example demonstrates the application of \pkg{asgl} to quantile regression, particularly in a scenario with a known group structure in the predictors, and showcases the effectiveness of the adaptive sparse group lasso (ASGL) penalization. We simulate data where variables are grouped, and only a subset of these groups is truly influential, a common situation in fields like genomics. The data generating function \code{make_group_data} can be found in the appendix. The training data will feature more predictors than samples ($p > n_{\text{train}}$). We compare \pkg{asgl}'s adaptive sparse group lasso quantile regression against a standard lasso-penalized quantile regression implemented in \pkg{scikit-learn}. Performance is evaluated based on predictive accuracy using the Median Absolute Error (MAE) and, critically, on the accuracy of coefficient estimation using the $\ell_2$ norm of the difference between the true and estimated coefficients ($\|\bm{\beta}_{\text{true}} - \hat{\bm{\beta}}\|_2$).

We generate a dataset with $n=200$ observations and $p=225$ features, structured into 15 groups of 15 variables each. Within each group, variables have a correlation of 0.5. Only 3 of these 15 groups are designed to be truly associated with the response variable. The data is split into training ($n_{\text{train}}=100$) and testing ($n_{\text{test}}=100$) sets. The \proglang{Python} code below details the full simulation, which iterates 25 times to produce robust performance metrics. The aggregated results are summarized in Table~\ref{tab:ex3_comparison}.

\begin{CodeInput}
>>> import time
>>> import numpy as np
>>> import pandas as pd
>>> from asgl import Regressor
>>> from sklearn.linear_model import QuantileRegressor
>>> from sklearn.model_selection import GridSearchCV, train_test_split
>>> from sklearn.metrics import median_absolute_error

# Simulation parameters
>>> n_repetitions = 25
>>> sklearn_lasso_grid = {'alpha': 10 ** np.arange(-2, 1, 0.1)}
>>> asgl_asgl_grid = {'lambda1': 10 ** np.arange(-2, 1, 0.1), 
...                   'alpha': np.arange(0, 1.01, 0.25)}

>>> results = []
>>> for it in range(n_repetitions):
...     print(f"Running iteration: {it+1}/{n_repetitions}")
...     X, y, beta_true, group_index = make_group_data(
...         n_obs=200, n_groups=15, group_size=15, corr=0.5, seed=it)
...     X_train, X_test, y_train, y_test = train_test_split(
...         X, y, test_size=100, random_state=it)

...     # 1. Lasso Quantile Regression using scikit-learn
...     start = time.perf_counter()
...     gscv_lasso_sk = GridSearchCV(
...         QuantileRegressor(), sklearn_lasso_grid, 
...         scoring='neg_median_absolute_error', 
...         cv=3, n_jobs=-1)
...     gscv_lasso_sk.fit(X_train, y_train)
...     lasso_sk_pred = gscv_lasso_sk.predict(X_test)
...     results.append({
...         'Model': 'scikit-learn lasso',
...         'MAE': median_absolute_error(lasso_sk_pred, y_test),
...         'L2 betas': np.linalg.norm(beta_true - 
...                         gscv_lasso_sk.best_estimator_.coef_),
...         'Exec time': time.perf_counter() - start})

...     # 2. Adaptive Sparse Group Lasso Quantile Regression using asgl
...     start = time.perf_counter()
...     gscv_asgl_asgl = GridSearchCV(
...         Regressor(model='qr', penalization='asgl'), 
...         asgl_asgl_grid, scoring='neg_median_absolute_error', 
...         cv=3, n_jobs=-1)
...     gscv_asgl_asgl.fit(X=X_train, y=y_train, group_index=group_index)
...     asgl_asgl_pred = gscv_asgl_asgl.predict(X_test)
...     results.append({
...         'Model': 'asgl adaptive sgl',
...         'MAE': median_absolute_error(asgl_asgl_pred, y_test),
...         'L2 betas': np.linalg.norm(beta_true - 
...                         gscv_asgl_asgl.best_estimator_.coef_),
...         'Exec time': time.perf_counter() - start})

# Results are then aggregated and summarized in Table 3
>>> df_results = pd.DataFrame(results)
>>> summary_ex3 = (df_ex3.groupby('Model')
...               [['MAE', 'L2 betas', 'Exec time']]
...               .agg(['mean', 'std']))
>>> print(summary)
\end{CodeInput}

\begin{table}[htb]
    \centering
    \caption{Example 3: Performance comparison for quantile regression in a high-dimensional setting with group structure ($n_{\text{train}}=100, p=225$, 15 groups of 15). Default $\tau=0.5$ (median regression) is used. Results are mean (standard deviation) over 25 iterations.}
    \label{tab:ex3_comparison}
    \begin{tabular}{lccc}
        \toprule
        \textbf{Model} & \textbf{MAE} & $\boldsymbol{\|\bm{\beta}_{\text{true}}-\hat{\bm{\beta}}\|_2}$ & \textbf{Execution time (s)} \\
        \midrule
        \pkg{scikit-learn} lasso Quantile Reg. & 12.89 (1.39) & 12.71 (2.09) & \textbf{0.21} (0.01) \\
        \pkg{asgl} Adaptive SGL Quantile Reg. & \textbf{11.30} (1.46) & \textbf{9.24} (3.68) & 5.87 (0.33) \\
        \bottomrule
    \end{tabular}
\end{table}

The performance metrics, averaged over 25 iterations, are presented in Table~\ref{tab:ex3_comparison}. These results highlight the benefits of employing the adaptive sparse group lasso (ASGL) penalization available in \pkg{asgl} for quantile regression, particularly in scenarios with grouped predictors. The \pkg{asgl} ASGL quantile regression model achieves a lower Median Absolute Error (MAE) of 11.30 compared to 12.89 for \pkg{scikit-learn}'s lasso-penalized quantile regression. More substantially, the \pkg{asgl} model demonstrates superior accuracy in coefficient estimation, with an $\ell_2$ norm of the coefficient error ($\|\bm{\beta}_{\text{true}}-\hat{\bm{\beta}}\|_2$) of 9.24, a marked improvement over the 12.71 achieved by the \pkg{scikit-learn} counterpart. This enhanced ability to recover coefficients closer to their true values is particularly valuable in applications where model interpretability and accurate identification of influential variables (and groups of variables) are paramount, such as in bioinformatics or econometrics.

Regarding computational cost, \pkg{scikit-learn}'s implementation is considerably faster (0.21 seconds) than \pkg{asgl}'s ASGL (5.87 seconds) for this specific configuration. This difference can be attributed to several factors: the ASGL model is inherently more complex, the hyperparameter search space for the \pkg{asgl} model in this example was larger (tuning both \code{lambda1} and \code{alpha}), and \pkg{asgl} relies on the general-purpose \pkg{cvxpy} backend, whereas \pkg{scikit-learn} often uses highly optimized, specialized solvers. The key advantage showcased here is that \pkg{asgl} makes sophisticated penalizations like ASGL readily available for complex models like quantile regression, empowering users to potentially achieve better model fit and more accurate parameter estimation, even if it involves a computational trade-off in some cases.

\subsection{Example 4: employing user-specified adaptive weights}

While \pkg{asgl} offers a comprehensive suite of built-in techniques for the automated estimation of adaptive weights (as detailed in Section~\ref{sec:adaptive_weights}), there are scenarios where users may possess specific statistical knowledge or have externally derived weights they wish to apply. Here, the package provides users the flexibility to employ their own pre-computed adaptive weights. This example demonstrates how to supply custom weights for adaptive penalizations, such as adaptive sparse group lasso (ASGL), using the parameters \code{individual_weights} (for weights $\tilde{w}_j$ applied to individual coefficients) and/or \code{group_weights} (for weights $\tilde{v}_l$ applied to groups) when instantiating the \class{Regressor}. When these parameters are provided, they override \pkg{asgl}'s internal weight estimation mechanisms (specified by \code{weight_technique}). The following code fits an ASGL model to synthetic high-dimensional data ($n=200, p=200$, with an arbitrary 5-group structure). For illustrative purposes, random weights are generated. In practice, user-supplied weights would be derived from a statistically sound basis. This feature grants users greater control in tailoring the penalization to their specific needs.

\begin{CodeInput}
>>> import numpy as np
>>> from sklearn.datasets import make_regression
>>> from asgl import Regressor

>>> np.random.seed(1234)

# Generate synthetic regression data
>>> X, y = make_regression(
...            n_samples=200, n_features=200, 
...            n_informative=25, bias=10, noise=5, random_state=42)

# Define the group structure
>>> group_index = np.random.randint(1, 5, size=200)

# Generate custom weights
>>> custom_individual_weights = np.random.rand(X.shape[1])
>>> custom_group_weights = np.random.rand(len(np.unique(group_index)))

# Create a Regressor object with custom weights
>>> model = Regressor(penalization="asgl", 
...            individual_weights=custom_individual_weights,
...            group_weights=custom_group_weights)

# Fit the model
>>> model.fit(X, y, group_index=group_index)

\end{CodeInput}

\subsection{Example 5: asgl vs scikit-learn performance comparison}\label{sec:example_5}

This example benchmarks the execution time of the lasso regression in the \pkg{asgl} package against the \code{Lasso} estimator from \pkg{scikit-learn}. For this example, the \pkg{asgl} package utilized the CLARABEL open-source solver, which is an interior-point solver for convex optimization problems. However, \pkg{asgl} can interface with various other open-source or commercial solvers 
. Using commercial solvers like Mosek or Gurobi may lead to further reductions in computation times. The complete \proglang{Python} code for this benchmark is provided in the appendix code file. The comparison involved two scenarios: first, a fixed number of observations ($n=200$) with a varying number of features ($p\in\left\{200, 500, 1000, 5000\right\}$), and second, a fixed $p=200$ with varying ($n\in\left\{200, 500, 1000, 5000\right\}$). For each configuration, datasets were generated 20 times with different seeds, and both models were fitted, with execution times recorded.

\begin{figure*}
	\centering
	\includegraphics[width=\textwidth]{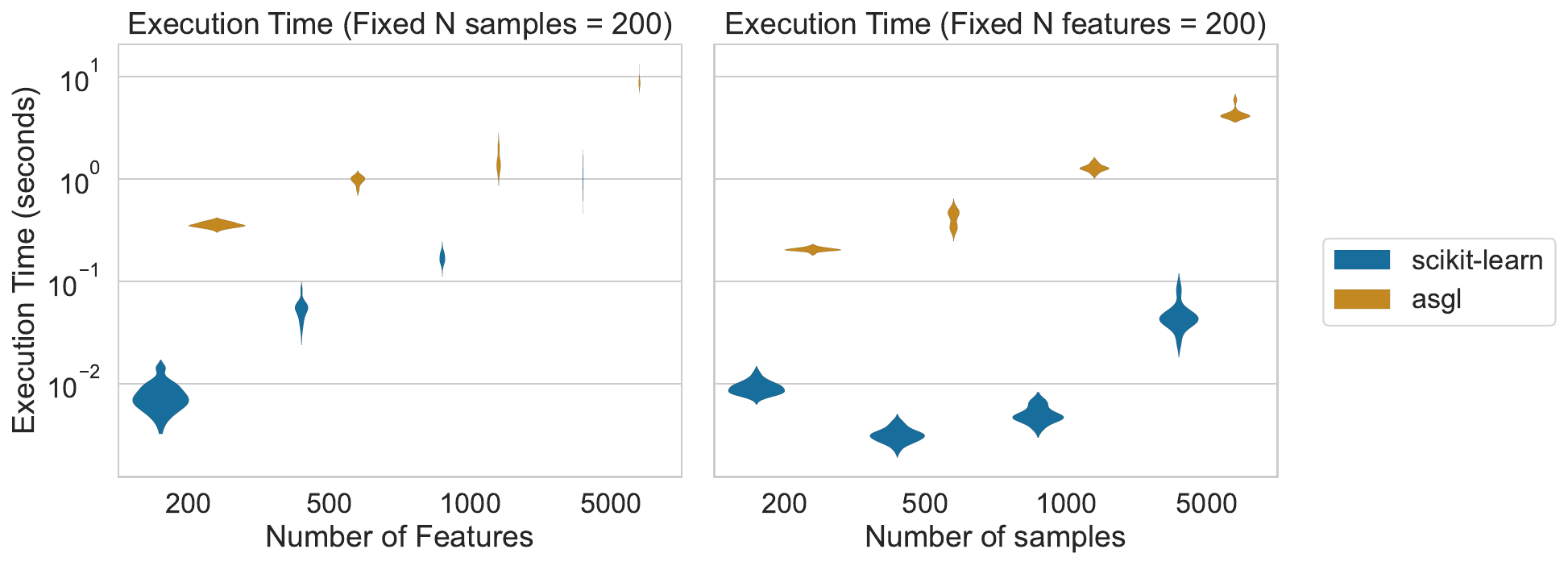}
	\caption{Execution times for scikit-learn's lasso and asgl's lasso algorithms. The left panel illustrates performance scaling with an increasing number of features while keeping the sample size fixed at 200. The right panel shows performance as the sample size increases with the number of features fixed at 200. Execution times derived from 20 repetitions per configuration}
	\label{fig:execution_times}
\end{figure*}

The results illustrated in Figure~\ref{fig:execution_times} show the violin plots of the execution times. These use kernel density plots where the width corresponds to the frequency of data points at different values. The results indicate that \pkg{scikit-learn} is generally faster. When increasing the number of features, both packages show computation times scaling logarithmically, though \pkg{scikit-learn} is consistently quicker (e.g., for $p=5000$, \pkg{asgl} averaged $\approx 6$s vs. $\approx 1.25$s for \pkg{scikit-learn}). With an increasing number of samples, \pkg{asgl} also scales logarithmically. In contrast, \pkg{scikit-learn} leverages highly optimized algorithmic implementations, resulting in lower computation times (e.g., for $n=5000$, \pkg{asgl} averaged $\approx 2.67$s vs. $\approx 0.05$s for \pkg{scikit-learn}). All computation times were obtained using the system described in Section~\ref{sec:computational_details}. Despite these speed differences for standard lasso, \pkg{asgl}'s execution times remain practical, particularly when considering its primary advantages in offering adaptive penalizations and penalized quantile regression and the improvement in performance metrics like mean squared error and in variable selection and estimation. 

\section{Computational details}\label{sec:computational_details}

All experiments were conducted with \pkg{asgl} package version 2.1.3 under Python 3.12 on a consumer laptop running Windows 11 Home. The machine is equipped with a 12th-generation Intel Core i7-12700H (20 logical cores at 2.3 GHz) and 32 GB RAM. On this hardware, examples 1–4  from Section~\ref{sec:examples} finish in roughly 12 s total. Example 5 adds $\approx$ 3 min, so the complete supplementary python script executes in under 4 min of wall-clock time. The package depends only on mainstream open-source libraries; the minimum tested versions are

\begin{itemize}
    \item \proglang{Python} >= 3.9
    \item \pkg{cvxpy} >= 1.5.0
    \item \pkg{numpy} >= 1.20.0
    \item \pkg{scikit-learn} >= 1.6
    \item \pkg{scipy} >= 1.1.0
    \item \pkg{pytest} >= 7.1.2 (primarily for development testing)
\end{itemize}

\pkg{asgl} is released under the GNU GPL v3, granting users the right to run, study, modify, and redistribute the software provided that any derivative remains under the same copyleft terms.

\section{Limitations}\label{sec:limitations}

While \pkg{asgl} offers a versatile framework for adaptive and penalized regression, some limitations should be noted. Computationally, \pkg{asgl}'s reliance on \pkg{cvxpy} for optimization, while ensuring flexibility, may result in longer execution times for standard tasks compared to packages with specialized solvers (as observed in Section \ref{sec:example_5} . Fitting complex models, such as adaptive sparse group lasso, or performing extensive hyperparameter tuning can be demanding. Scalability to datasets substantially larger than those tested (e.g., hundreds of thousands of features/observations) has not been exhaustively evaluated and might be constrained by the general-purpose backend. The current scope of \pkg{asgl} encompasses linear, logistic, and quantile regression. Future extensions could incorporate other model families (e.g., other GLMs, survival models etc.). The package focuses on coefficient estimation and prediction and does not currently implement specialized methods for post-selection inference (e.g., p-values or confidence intervals for selected coefficients). For group penalizations, the group structure (\code{group_index}) must be pre-specified by the user, as \pkg{asgl} does not offer automated group discovery. Regarding adaptive weight estimation, while \pkg{asgl} provides a valuable suite of automated techniques, the optimal choice of \code{weight_technique} and its associated hyperparameters can be data-dependent, potentially requiring careful cross-validation. Despite these limitations, \pkg{asgl} provides a powerful and accessible tool for a wide array of modern regression problems, particularly excelling in its unified support for adaptive penalizations and penalized quantile regression.


\section{Conclusion}\label{sec:conclusions}

The \pkg{asgl} \proglang{Python} package has been introduced as a comprehensive and user-friendly framework for fitting linear, logistic, and quantile regression models with a wide array of penalization techniques. This paper has detailed its theoretical underpinnings, software architecture centered around the \pkg{scikit-learn} compatible \class{Regressor} class, and practical application through illustrative examples. \pkg{asgl} particularly distinguishes itself by its extensive support for adaptive penalizations—including adaptive lasso, adaptive ridge, adaptive group lasso, and adaptive sparse group lasso—across all three model types.

A core contribution of \pkg{asgl} is its robust implementation of multiple built-in methodologies for the automatic estimation of adaptive weights (e.g., using PCA, PLS, ridge, or lasso initial estimates). This critically addresses a common practical hurdle in applying adaptive methods, especially in high-dimensional settings where obtaining reliable initial estimates for weights is challenging. By automating weight calculation, \pkg{asgl} makes the well-documented theoretical benefits of adaptive penalizations, such as reduced bias and improved variable selection consistency, practically accessible to a broader audience, alleviating concerns about the complexity of determining appropriate weight vectors.

Compared to existing alternatives, \pkg{asgl} offers a unique combination of versatility and advanced features. While packages like \pkg{scikit-learn} provide excellent support for standard penalizations such as lasso and ridge, \pkg{asgl} extends this by offering a richer set of penalizations (e.g., group lasso, sparse group lasso) and, most importantly, their adaptive variants with integrated weight estimation—a feature largely absent or requiring manual intervention in other \proglang{Python} tools or even established \proglang{R} packages like \pkg{glmnet}. Furthermore, \pkg{asgl}'s robust support for penalized quantile regression, including adaptive versions, fills a significant niche. The seamless integration with the \pkg{scikit-learn} API ensures that users can easily incorporate \pkg{asgl} into their existing workflows, leveraging familiar tools for hyperparameter tuning and model evaluation. Indeed, existing \pkg{scikit-learn} pipelines can often be augmented by switching to \pkg{asgl} as an almost drop-in replacement, with the potential for enhanced predictive performance and more accurate variable selection, as demonstrated in our examples. By reducing the implementation burden and promoting reproducible research, \pkg{asgl} empowers both researchers and practitioners in various fields to apply state-of-the-art penalized and adaptive regression methods to their data.

Future development of \pkg{asgl} will focus on expanding its capabilities, potentially including support for additional model families (such as other GLMs or survival models), exploring further computational optimizations, and incorporating methods for post-selection inference. We believe that \pkg{asgl} serves as a valuable addition to the statistical software landscape, facilitating the broader application of advanced regression modeling techniques and contributing to more insightful data analysis.

\section{Acknowledgements}

This work was partially supported by the Spanish Ministry of Science and Innovation \\(MCIN/AEI/10.13039/501100011033) under grant PID2022-137243OB-I00. This work also received support from the European Union’s Recovery, Transformation and Resilience Plan – NextGenerationEU, through the INCIBE ANTICIPA grant and the ENIA 2022 programme for university–industry AI chairs (AImpulsa: UC3M-Universia).


\bibliography{references}

@article{scikit-learn,
	title = {Scikit-learn: {Machine} learning in {Python}},
	volume = {12},
	journal = {Journal of Machine Learning Research},
	author = {Pedregosa, F. and Varoquaux, G. and Gramfort, A. and Michel, V. and Thirion, B. and Grisel, O. and Blondel, M. and Prettenhofer, P. and Weiss, R. and Dubourg, V. and Vanderplas, J. and Passos, A. and Cournapeau, D. and Brucher, M. and Perrot, M. and Duchesnay, E.},
	year = {2011},
	pages = {2825--2830},
}

@misc{friedman_note_2010,
	title = {A note on the group lasso and a sparse group lasso},
	url = {http://arxiv.org/abs/1001.0736},
	doi = {10.48550/arXiv.1001.0736},
	urldate = {2024-09-12},
	publisher = {arXiv},
	author = {Friedman, J. and Hastie, T. and Tibshirani, R.},
	month = jan,
	year = {2010},
	note = {arXiv:1001.0736 [math, stat]},
}

@article{liang_sparsegl_2024,
	title = {sparsegl: {An} {R} {Package} for {Estimating} {Sparse} {Group} {Lasso}},
	volume = {110},
	copyright = {Copyright (c) 2024 Xiaoxuan Liang, Aaron Cohen, Anibal Sólon Heinsfeld, Franco Pestilli, Daniel J. McDonald},
	issn = {1548-7660},
	shorttitle = {sparsegl},
	url = {https://doi.org/10.18637/jss.v110.i06},
	doi = {10.18637/jss.v110.i06},
	language = {en},
	urldate = {2024-09-11},
	journal = {Journal of Statistical Software},
	author = {Liang, Xiaoxuan and Cohen, Aaron and Heinsfeld, Anibal Sólon and Pestilli, Franco and McDonald, Daniel J.},
	month = aug,
	year = {2024},
	pages = {1--23},
}

@article{hoerl_ridge_1970,
	title = {Ridge {Regression}: {Biased} {Estimation} for {Nonorthogonal} {Problems}},
	volume = {12},
	issn = {0040-1706},
	shorttitle = {Ridge {Regression}},
	url = {https://www.jstor.org/stable/1267351},
	doi = {10.2307/1267351},
	number = {1},
	urldate = {2024-09-11},
	journal = {Technometrics},
	author = {Hoerl, Arthur E. and Kennard, Robert W.},
	year = {1970},
	note = {Publisher: [Taylor \& Francis, Ltd., American Statistical Association, American Society for Quality]},
	pages = {55--67},
}

@article{simon_sparse-group_2013,
	title = {A {Sparse}-{Group} {Lasso}},
	volume = {22},
	issn = {1061-8600, 1537-2715},
	url = {http://www.tandfonline.com/doi/abs/10.1080/10618600.2012.681250},
	doi = {10.1080/10618600.2012.681250},
	language = {en},
	number = {2},
	urldate = {2024-08-20},
	journal = {Journal of Computational and Graphical Statistics},
	author = {Simon, Noah and Friedman, Jerome and Hastie, Trevor and Tibshirani, Robert},
	month = apr,
	year = {2013},
	pages = {231--245},
}

@article{he_smoothed_2023,
	title = {Smoothed quantile regression with large-scale inference},
	volume = {232},
	issn = {0304-4076},
	url = {https://www.sciencedirect.com/science/article/pii/S0304407621001950},
	doi = {10.1016/j.jeconom.2021.07.010},
	number = {2},
	urldate = {2024-04-29},
	journal = {Journal of Econometrics},
	author = {He, Xuming and Pan, Xiaoou and Tan, Kean Ming and Zhou, Wen-Xin},
	month = feb,
	year = {2023},
	pages = {367--388},
}

@article{Diamond2016,
	title = {{CVXPY}: {A} {Python}-{Embedded} {Modeling} {Language} for {Convex} {Optimization}},
	urldate = {2019-03-15},
	journal = {arXiv:1603.00943},
	author = {Diamond, Steven and Boyd, Stephen},
	month = mar,
	year = {2016},
	note = {arXiv: 1603.00943},
}

@article{Wright2010,
	title = {Sparse {Representation} for {Computer} {Vision} and {Pattern} {Recognition}},
	volume = {98},
	issn = {0018-9219},
	doi = {10.1109/JPROC.2010.2044470},
	number = {6},
	urldate = {2019-03-15},
	journal = {Proceedings of the IEEE},
	author = {Wright, John and Ma, Yi and Mairal, Julien and Sapiro, Guillermo and Huang, Thomas S. and Yan, Shuicheng},
	month = jun,
	year = {2010},
	note = {1371 citations (Crossref) [2023-10-13]},
	pages = {1031--1044},
}

@article{Yuan2006,
	title = {Model selection and estimation in regression with grouped variables},
	volume = {68},
	doi = {10.1111/j.1467-9868.2005.00532.x},
	number = {1},
	urldate = {2019-03-15},
	journal = {Journal of the Royal Statistical Society. Series B (Methodological)},
	author = {Yuan, Ming and Lin, Yi},
	year = {2006},
	pages = {49--67},
}

@article{Zou2005,
	title = {Regularization and variable selection via the elastic net},
	volume = {67},
	doi = {10.1111/j.1467-9868.2005.00503.x},
	number = {2},
	urldate = {2019-03-15},
	journal = {J. R. Statist. Soc. B},
	author = {Zou, Hui and Hastie, Trevor},
	year = {2005},
	pages = {301--320},
}

@article{Tibshirani1996,
	title = {Regression {Shrinkage} and {Selection} via the {Lasso}},
	volume = {58},
	doi = {10.2307/2346178},
	number = {1},
	urldate = {2019-03-15},
	journal = {Journal of the Royal Statistical Society. Series B (Methodological)},
	author = {Tibshirani, Robert},
	year = {1996},
	note = {Publisher: WileyRoyal Statistical Society},
	pages = {267--288},
}

@article{Mendez-Civieta2020,
	title = {Adaptive sparse group {LASSO} in quantile regression},
	volume = {15},
	copyright = {All rights reserved},
	issn = {18625355},
	doi = {10.1007/s11634-020-00413-8},
	number = {3},
	journal = {Advances in Data Analysis and Classification},
	author = {Mendez-Civieta, Alvaro and Aguilera-Morillo, M. Carmen and Lillo, Rosa E.},
	year = {2021},
	note = {4 citations (Crossref) [2023-10-13]
Publisher: Springer Berlin Heidelberg},
	pages = {547--573},
}

@article{Rapach2013,
	title = {International {Stock} {Return} {Predictability} : {What} {Is} the {Role} of the {United} {States} ?},
	volume = {68},
	doi = {10.1111/jofi.12041},
	number = {4},
	journal = {The Journal of Finance},
	author = {Rapach, David E and Strauss, Jack K and Zhou, Guofu},
	year = {2013},
	note = {485 citations (Crossref) [2023-10-13]},
	pages = {1633--1662},
}

@article{Zou2006,
	title = {Sparse {Principal} {Component} {Analysis}},
	volume = {15},
	doi = {10.1198/106186006X113430},
	number = {2},
	urldate = {2019-05-15},
	journal = {Journal of Computational and Graphical Statistics},
	author = {Zou, Hui and Hastie, Trevor and Tibshirani, Robert},
	year = {2006},
	note = {1809 citations (Crossref) [2023-10-13]},
	pages = {265--286},
}

@article{Simon2013,
	title = {A sparse-group lasso},
	volume = {22},
	issn = {10618600},
	doi = {10.1080/10618600.2012.681250},
	number = {2},
	urldate = {2019-03-15},
	journal = {Journal of Computational and Graphical Statistics},
	author = {Simon, Noah and Friedman, Jerome and Hastie, Trevor and Tibshirani, Robert},
	month = apr,
	year = {2013},
	note = {811 citations (Crossref) [2023-10-13]
Publisher: Taylor \& Francis Group},
	pages = {231--245},
}

@article{Zou2006a,
	title = {The {Adaptive} {Lasso} and {Its} {Oracle} {Properties}},
	volume = {101},
	issn = {0162-1459},
	doi = {10.1198/016214506000000735},
	number = {476},
	urldate = {2019-03-15},
	journal = {Journal of the American Statistical Association},
	author = {Zou, Hui},
	month = dec,
	year = {2006},
	note = {4325 citations (Crossref) [2023-10-13]
Publisher: Taylor \& Francis},
	pages = {1418--1429},
}

@article{Koenker1978,
	title = {Regression {Quantiles}},
	volume = {46},
	issn = {00129682},
	doi = {10.2307/1913643},
	number = {1},
	urldate = {2019-03-15},
	journal = {Econometrica},
	author = {Koenker, Roger and Bassett, Gilbert},
	month = jan,
	year = {1978},
	note = {8985 citations (Crossref) [2023-10-13]
Publisher: The Econometric Society},
	pages = {33--50},
}

@article{Friedman2010b,
	title = {Regularization {Paths} for {Generalized} {Linear} {Models} via {Coordinate} {Descent}.},
	volume = {33},
	number = {1},
	urldate = {2019-03-15},
	journal = {Journal of statistical software},
	author = {Friedman, Jerome and Hastie, Trevor and Tibshirani, Rob},
	year = {2010},
	pmid = {20808728},
	pages = {1--22},
}

\end{document}